\documentclass[runningheads]{svjour2}
\smartqed  % flush right qed marks, e.g. at end of proof
\usepackage{graphicx}
\usepackage{biograph}        % to allow for author biography at the end
\newcommand{\beq}{\begin{equation}}                       % Abbreviare
 \newcommand{\eeq}{\end{equation}}  
\newcommand{\vs}{\vspace{3mm}} 

\newcommand{\no}{\noindent}

\journalname{Ricerche di Matematica}
\begin{document}

\title{Existence, uniqueness and a priori estimates for a non linear integro - differential equation%\thanks{Grants or other notes
}

%\titlerunning{Short form of title} % if too long for running head


%\dedication{Short devotement}      % optional dedication of article

\author{M. De Angelis         \and
        P. Renno %etc.
}


\institute{M.De Angelis \at
             University of Naples Federico II,Faculty of Engineering. Dept. of Math. and Appl. \\ Via Claudio 21,  80125 Naples,  Italy \\
              \email{modeange@unina.it }           %  \\
           \and
           P. Renno\at
              University of Naples Federico II. Faculty of Engineering. Dept. of Math. and Appl. \\ Via Claudio 21,  80125 Naples,  Italy \\
              \email{renno@unina.it } 
}
\date{Received: November 21, 2007/ Accepted: December 28, 2007}

\maketitle

\begin{abstract}
The paper deals with the explicit calculus and the properties of the fundamental solution $\, K\,$ of a parabolic operator related to a semilinear equation that models reaction diffusion systems with excitable kinetics. The initial value problem in all of the space is analyzed together with continuous dependence and a priori estimates of the solution. These estimates show that the asymptotic behavior is determined by the reaction mechanism. Moreover it's possible a rigorous singular perturbation analysis for discussing travelling waves with their characteristic times.
\end{abstract}

\section{Statement  of the problem and results}
\label{intro}
Let $\,   \Omega_T \, \equiv \{\,(x,t) :  x \in \Re \,\,  \ 0 < t \leq T \, \}$ and, for functions  $\, u(x,t) \in  C^2 (\,\Omega_T \,),\, $ let  $\,{\cal P}_0\,$  be the non  linear initial value problem defined in all of the space:

	  \beq                                                     \label{11}
  \left \{
   \begin{array}{lll}
   Lu\, \equiv \,  u_t -  \varepsilon  u_{xx} + au +b \int^t_0  e^{- \beta (t-\tau)}\, u(x,\tau) \, d\tau \,=\, F(x,t,u) \,\\    
\\
  \,u (x,0)\, = g(x)\,  \,\,\,\,\, x\, \in \, \Re, 
\\
   
   \end{array}
  \right.
 \eeq

\no where  $\, a,\,b,\, \varepsilon, \, \beta \, $ are positive constants and $ \, F,\, g\, $ are known functions of their arguments.

If $\, K(x,t) \, $ is  a fundamental solution of the parabolic operator $\, L\, $  and $\, F(\,x,\,t,\,u\,)\,$ verifies appropriate assumptions, then the differential problem  (\ref{11}) is equivalent to the integral equation

%\vspace{3mm}
\beq                       \label{12}
u(x,t) \,\, =\, \,{\cal F}\,[ \, u(x,t)\,],
\eeq
   
 \no where $\,\,{\cal F}\,[ \, v\,]$ is the mapping

  \beq                        \label{13}
   {\cal F}\,[\,v\, (x,t)\,] \,\, =\,\,\int_\Re   \,K ( x-\xi, t)\,\, \,g (\xi)\,\,d\xi \,\,  + \,\,
\eeq
\\
\[\,+\,\int ^t_0     d\tau \int_\Re   K ( x-\xi, t-\tau)\,\, F\,[\,\xi,\tau, v(\xi,\tau\,)\,]\,\, d\xi. \]

\vs\vs
 Now, let  $\,\,|| \, v \, ||_T \,  = \displaystyle{\sup_{\Omega_T}}\,| v (x,t)\,|, \,\,$  and let  $ \,{\cal B}_ T \,$  denote the Banach space

\vs\beq     \label{14}
  \,{\cal B}_ T \, \equiv \, \{\, v\,(\,x,t\,) : \, v\, \in  C (\,\Omega_T \,),\,  ||v||_T \, < \infty \ \}.
\eeq

\vspace{1mm}The aim of the paper  is the explicit calculus of  $\, K\,$ and the analysis of its basic properties (\S 2)  useful to prove that  	$\,{\cal F}\,$ is a contraction of $\,{\cal B}_ T \,$ in $\,\,{\cal B}_ T \,$ and hence admits a unique fixed point $\, u\,=\, u (x,t)\, \in \,{\cal B}_ T. \, \,$ Thus existence and uniqueness results for the problem $\,{\cal P}_0\,$ are deduced (\S 3), together with continuous dependence and a priori estimates of the solution  (\S 4). 

\vs The following estimate (\S 2) is worthy of remark

\beq               \label{15}
|K(x,t)| \, \leq \, \frac{e^{- \frac{x^2}{4 \varepsilon \,t}\,}}{2\,\sqrt{\pi \varepsilon t}} \,\, [ \, e^{\,-\,at}\, +\, b t \,\,\,\frac{e^{\,-\,a t}\,-\,e^{\,-\beta\,t}}{\,\beta\,-\,a\,}\,\, ] 
\eeq

\vs \no and it shows that $\, K\,$ has the same basic properties of the fundamental solution of the heat operator and  moreover decays exponentially to zero as $\,t\,$ increases. As consequence (\S \, 4) , the solution $\,u\,$ of $\,{\cal P}_0\,$ is such that

\vspace{3mm}
\beq               \label{16}
||u(x,t)||_T \,\, \leq \,\,||g ||\,\,( 1\,+\,\,\pi \,  \sqrt b\,\,t\,) \,\,\, e^{\,-\omega \,t\,} \, +\,\, \beta_0 \,\,\,||F ||,\,\, \,
\eeq

%\vspace{3mm}
%\beq                               \label{15}
%|K(x,t) |\, \leq \,\,\, ( 1+t \, \sqrt b\,) \,\,\, e^{\,-\,k\,t\,} \,\,\frac{ \, e^{-\, \frac{x^2}{4\varepsilon \,t }\,}}{ \sqrt{\pi \varepsilon \,t}}\,\,\eeq

\vs \no where $\, \omega\,=\, \, \min \,(a,\beta\,),\,$ the norms are defined in \S 4 and the constant $\,\beta_0 $ depends on $\, a\,, \,b,\, \beta\,$ (see (\ref{225})\,). So, when $\,t \, $ is large, the asymptotic behavior of $\,u\,$ is determined by the properties of the source $\,F.\,$

 The equation (\ref{11}) models   several engineering  applications as  motions of viscoelastic fluids or solids  \cite{bcf,dr1,jp,r}; heat conduction at low temperature \cite{jcl,mps}, sound propagation in viscous gases \cite{l} and  perturbed sine Gordon equation in the theory of the superconductivity  \cite{bp,dm,acscott}  or in the propagation of localized magnetohydrodinamic models in plasma physics \cite{sbec}.

 As an example, in section 5 the results are applied to the  FitzHugh - Nagumo equations (FHN) which model many important biological  phenomena \cite{aa,i,k,ks,ri}. According to well known results concerning systems of non linear reaction- diffusion equations \cite{m,sm}, estimates like (\ref{16}) confirm that the large time behaviour of the solution is determined by the reaction mechanism. Moreover by means of the explicit fundamental solution $\,K\,$ and its properties, it's possible  a rigorous singular perturbation analysis to approximate travelling pulses in excitable media, together with their transition times.

%Your text comes here. Separate text sections with

\section{Fundamental solution and properties}
\label{sec:2}
By denoting with

\[\hat u (x,s) \, = \int_ 0^\infty \, e^{-st} \, u(x,t) \,dt \,\,, \,\,\,  \,\hat F (x,s)   \, = \int_ 0^\infty \,\, e^{-st} \,\, F\,[x,t,u (x,t)\,] \,dt \,,\,\]

\noindent the Laplace transform with respect to $\,t,\,$ from (\ref{11}) we get

  \beq                                                     \label{21}
  %\left \{
   %\begin{array}{lll}
%\begin{center}
   {\hat u }(x,s)\,=\,  \int_\Re \, \hat K \,(\, x-\xi, s\,)\, [\,g(\,\xi\,) \,+\,\hat F(\xi,s)\,]\,d\xi\,, 
%\end{center}
\eeq

\noindent where

\beq                 \label{22}
\hat K (x,s) = \,\, \frac{e^{- \frac{|x|}{\sqrt \varepsilon} \,\,\sigma}}{2 \,\, \sqrt\varepsilon \,\,\,\sigma  }  \ \ \  with \,\,\,\,\,\, \sigma^2 \ \,=\, s\, +\, a \, + \, \frac{b}{s+\beta}.\,\,
\eeq

\vs Therefore, if $\, K (x,t)\,$ represents   the inverse $ {\cal L}_t$ transforms    of  $\, \hat K ( x,s),\, $ from (\ref{21}) {\em  formally}   it follows that

\vs  \beq                        \label{23}
   u (x,t)  =\int_\Re   \,K ( x-\xi, t)\,\, g (\xi)\,\,d\xi \,\,  + \,\,
\eeq
\\
\[\,+\,\int ^t_0     d\tau \int_\Re   K ( x-\xi, t-\tau)\,\, F\,[\,\xi,\tau, u(\xi,\tau\,)\,]\,\, d\xi. \]

 \vs If $\, r\,= |x| \, / \sqrt \varepsilon \, \, $ and $ J_n (z) \,$    denotes the Bessel function of first kind and order $\, n,\,$ let us  consider the function

   \beq     \label {24}
K(r,t)\,=\, \, \, \frac{e^{- \frac{r^2}{4 t}\,}}{2 \sqrt{\pi  \varepsilon t } }\,\,\, e^{-\,a\,t}\,-\,
\eeq
\\
 \[ \, - \,\,  \frac{1}{2} \,\, \sqrt{ \frac{\,b}{\pi \, \varepsilon \,}}   \,\,\,\int^t_0  \frac{e^{- \frac{r^2}{4 y}\,- a\,y}}{\sqrt{t-y}} \,\, \,\ e^{-\beta \,(\, t \,-\,y\,)} \, J_1 (\,2 \,\sqrt{\,b\,y\,(t-y)\,}\,\,)\,\,dy.\,\,
 \]

\vs{\bf Theorem 2.1}- {\em In the half-plane} $ \Re e  \,s > \,max(\,-\,a ,\,-\beta\,)\,$   {\em the Laplace integral}  $\,{\cal L  }_t\,\,K (r,t)\, \,$ {\em  converges absolutely for all  }$\,r>0,\,$  {\em and it results}:

\vs
\beq      \label{25}
\,{\cal L  }_t\,\,K\,\equiv \,\,\int_ 0^\infty e^{-st} \,\, K\,(r,t) \,\,dt \,\,=  \, \frac{e^{- \,r\,\sigma}}{2 \, \sqrt\varepsilon \,\sigma \,  }.
\eeq

 \vs
{\bf Proof -} For all real $\, z,\,$ one has $\, |J_n\,(z)|\, \leq \, 1  $ and the Fubini -Tonelli theorem implies that

 \vs         
\[
\,\,{\cal L  }_t \,K \, =\, \frac{e^{\,-r\, \sqrt{\,s+a\,}}}{2\, \sqrt{\,\varepsilon \,(\, s+a)\,}}\,\,- \]
\\
  \[
\, - \,\, {\frac{ \sqrt b }{2\, \sqrt{\pi  \varepsilon  }} }\, \,\,\,\ \int ^\infty_0   e^{- (s+a)y \,-\frac{ r^2}{4 y} } 
\,\,dy\,\,\int ^\infty_0 e^{-(s+\beta)t}\, \,\,J_1 (2 \sqrt{b \,y\,t  }\,) \,\,\frac{dt}{\sqrt t}.\,\, 
\]
 %\end{center}

\vs Since

\begin{center}   
%\beq 
\[
%\begin {aligned}
 \,\, \int ^\infty_0    \,\, {e^{- p\, t} }\,\,  \sqrt{\frac{c}{ t}}\,\,\,\,J_1 ( \, 2 \sqrt{c\,t}\,\,)\,\,dt \,=\, \, 1\,-\, e^{-\, c/ p\,}\,\,\,\,\,\,\, (\,\Re e  \,p > \,0\,), \]
%\eeq

\end{center}
\vs\noindent it follows that

 \vs
\[
\,  \hat K \, (r,s)\,=\, \frac{1}{2\, \sqrt{\pi \varepsilon }}\,\int_ 0^\infty \,\,e^{\,-\,\frac{\,r^2\,}{4y}\,-\, (s+a+\frac{\,b\,}{s+\beta\,}\,)\,y\,}\,\,\frac{dy\,}{\sqrt y\,}\,\,= \,\, \frac{1}{2 \,\, \sqrt\varepsilon \, \,}\,\,\,\frac{e^{- \,r\,\sigma}}{\,\sigma \,}.\,\,\,\,\,\,\,\,\,\ \hbox{}\hfill\rule{1.85mm}{2.82mm} \]

\vspace{4mm}{\bf Theorem 2.2} - {\em The function}  $\  K \, $ {\em  has the same  basic properties of the fundamental solution of the heat equation, that is }:

%\beq                 \label{46} 

\vs A - \hspace{2mm} $ \,\,K(x,t) \, \, \in  C ^ {\infty} \,\,\,\,${\em for} \,$\,\,\, t>0, \,\,\,\, x \,\,\, \in \Re. $ 
%\eeq

\vs B - \hspace{2mm} {\em For fixed }$\, t\,>\,0,\,\,\, K \,$ {\em and its derivatives are vanishing esponentially fast as} $\, |x| \, $ {\em tends to infinity}.

\vs C - \hspace{2mm} {\em  For any fixed }$\, \delta \,>\, 0,\, $ {\em uniformly for all} $\, |x| \,\geq \, \delta,\, ${\em it results}:

\beq                    \label{26}
 \lim _{t\, \downarrow 0}\,\,K(x,t)\,=\,0,
\eeq

\vs D. - \hspace{2mm} {\em For }$\,t\,>\,0,\,$  {\em it is } $\,\,\, L\,K    =\, 0.  \,\,$

\vs{\bf Proof -} By (\ref{24}) the properties A-B-C are obvious. To verify D we put

\beq                           \label{27}
\psi(r,t)\,=\,  \, \,  \frac{1}{{2 \sqrt{\pi  \varepsilon t } }\,\,}\,\,\exp {\, [ \,- \frac{r^2}{4 t}\,-\,a\,t\,]\,}\,
\eeq
 
\beq                           \label{28}
\varphi\,(\,y,t\ )\,=\, \sqrt{ \frac{by}{t-y }\,\,} \,\  \, J_1 (\,2 \,\sqrt{\,b\,y\,(t-y) \,}\,\,)\,\,e^{-\beta \,(\, t \,-\,y\,)}
\eeq

\beq                             \label{29}
K(r,t)\,=\, \psi(r,t)\, -\, \int_0^t \, \, \psi  \,(r,y)\, \, \varphi(y,t)\, dy.
\eeq

\vspace{3mm}
As

\beq                              \label{210}
 (\, \partial_t\,+\, a\, -\, \partial_{rr}\,)\,\ \psi  \,(r,t)\,= \,0
\eeq

 \noindent  from (\ref{29}) it results:

\[
%\begin{aligned}
 K_t+ aK - K_{rr}=\, - \varphi (t,t)\, \psi  \,(r,t)- \int_0^t  [ \psi  ( \varphi_t +a\varphi) - \varphi \psi_{rr}] dy =
\]

\beq                              \label{211}
\,=\, -\varphi \,(t,t)\, \psi  \,(r,t)- \int_0^t  [ \psi  ( \varphi_t +a\varphi) -\, \varphi ( \psi_y\,+\,a\psi\,)\,] \,dy .
\eeq

 \noindent But

%\beq                              \label{49}

\[ \int_0^t  \,\,  \varphi \, \psi_y \, dy\, =  \varphi \,(t,t)\,\,\, \psi  \,(r,t)- \int_0^t \,\,  \psi \,\, \varphi_y \,\, dy \]

%\eeq

\vs\noindent and so (\ref{211}) gives

\beq                              \label{212}
 K_t+ aK - K_{rr}\,\,=\, -\,  \int_0^t  \,\, \psi  \,\,( \,\varphi_t \, +\,\varphi_y \, ) \, dy. 
\eeq

  \noindent As

\beq                              \label{213}
 (\, \partial_t +\, \partial_y \,) \,\varphi \, (y,t)  \,= \,b\, \,\,J_0 (2 \sqrt{b \,y\, (t-y)  }\,) \,\, e^{-\, \beta (\, t-y)\,} 
\eeq

 \vs\noindent from (\ref{212}) it follows
\vs
\beq                              \label{214}
 K_t+ aK - K_{rr}\,\,=\, -\, b\,\,K_1
\eeq

 \noindent where $\, K_1\, $ is given by 

\beq     \label {215}
K_1( r,t) \, = \, \frac{1}{2 \, \sqrt{\pi \, \varepsilon}} \,\, \int^t_0 \,\, e^{- \frac{r^2}{4 y}\,\,-a \,y\,-\,\beta (t-y)}\ \, \, J_0 \,(2 \,\sqrt{b\,y\,(t-y)}\,)\,\,\frac{dy}{\sqrt{y}}.
\eeq

\vs On the other side, the convolution $\, e^{-\,\beta \,t} \,*\,K \,$ is

\beq                              \label{216}
  e^{-\,\beta \,t} *K = e^{-\beta \,t} *\psi \,- \, \int^t_0 \, \psi \,(r,y)\, dy \, \int^t_y \, e^{-\,\beta \,(t-\tau)\,} \varphi \,(y,\tau)\, d\tau 
\eeq

 \noindent and moreover it results:

\beq                              \label{217}
  \int^t_y    e^{-\beta (t-\tau)}  \varphi ( y,\tau) d\tau=   e^{-\beta (t-y)}  \int ^t_y \sqrt{\frac{{by}}{\tau -y } }\,\,\,J_1 \, (\, 2\, \sqrt{b\, y ( \tau-y)  }\,)\,\, d\tau =
\eeq
\[ \,=\,\, e^{\,-\, \beta \,(t-y)} \,  \ 
\Bigl[
\,1\,-\ \,J_0 \, (\, 2\, \sqrt{b\, y ( t-y)  }\,)\,\,
\Bigr]. \]

\vs \noindent As consequence, from (\ref{216}) - (\ref{217})  we get

\beq                              \label{218}
  e^{-\,\beta \,t} *\,K =  \, \int^t_0 \, e^{-\beta \,(t- y)}\, \psi \,(r,y)\, \, J_0 \, (\, 2\, \sqrt{b\, y ( t-y)  }\,) \,\,dy = K_1
\eeq

%\[  J_0 \,\biggl(\, 2\,\,\,\root \of{\,  {b\, y ( t-y)  }\,}\,\, \biggr)\]

\noindent  and so  (\ref{214}) - (\ref{218}) imply property D because $\, K_{rr}\, = \,\varepsilon \, K_{xx}. $\hbox{}\hfill\rule{1.85mm}{2.82mm}

 \vspace{5mm}  Moreover, if $ \, I_n (z) $ denotes the modified Bessel function of the real variable $\, z\,$ and

\beq      \label{219} 
E(t) \,=\, \frac{e^{\,-\,\beta t}\,-\,e^{\,-\,at}}{a\,-\,\beta}\,\,>0\,,
\eeq

\vs\no  then the following theorem holds:

\vspace{4mm} {\bf Theorem 2.3} - {\em The fundamental solution }  $\  K \,(r,t) $  {\em satisfies the estimates}

\beq               \label{220}
|K| \, \leq \, \frac{e^{- \frac{r^2}{4 t}\,}}{2\,\sqrt{\pi \varepsilon t}} \,\, [ \, e^{\,-\,at}\, +\, b t \,E(t)\, ] 
\eeq

\beq               \label{221}
\int_\Re\,\,|K(x-\xi,t)|\,d\xi\,\,\leq \, e^{\,-\,at}\, +\, \sqrt b\, \pi \,t \,  \, e^{\,-\,\frac{\beta +a }{2}\,t} \,\, I_0 \biggl( \frac{\beta -a }{2}\,t \biggr).
\eeq

\vspace{5mm} {\bf Proof -} As $\, |J_1 ( \,2 \, \sqrt{by (t-y)}\,|\, \leq \sqrt{by (t-y)}, \,$ from  (\ref{24}) we get  (\ref{220}). Further, because

\[               
\int_\Re  \displaystyle {\frac{e^{- \frac{|x-\xi|^2}{4 \varepsilon t}\,}}{2\,\sqrt{\pi \varepsilon t}}} \, d\xi \, = \, 1, 
\]

\vs\no from (\ref{24}) and $ \, | J_1 (z) | \, \,\leq 1 \,$ it results:

\[ \int _\Re \, |K(x-\xi,t)|\,d\xi\,\,\leq \, e^{\,-\,at}\, +\int_0^t \, e^{\,- ay\,-\beta (t-y) \,} \, \sqrt{\frac{by}{t-y}}\,dy\,= \]

\[ =  e^{-at}\,+\,( \sqrt{\,b}\,/\,2\,) \,\pi \,t \, e^{\,-\,\frac{\beta +a }{2}\,t} \,\, \biggl[\,I_0 \left( \frac{a-\beta  }{2}\,t\right) -\, I_1 \left( \frac{a-\beta  }{2}\,t\right) \,\biggr] .\]

 \noindent As  $  \, I_1 (-z) = - I_1(z), \,\,I_0 (-z) =  I_0(z) $ and $   \, I_1(|z|) < I_0 (|z|), $  the estimate  (\ref{221}) follows.  \hbox{}\hfill\rule{1.85mm}{2.82mm}

\vspace{7mm} Now, if one puts $ \, \omega = \min \,(a, \beta ), $ by means of  (\ref{221}) we get 

\vs
\beq               \label{223}
\int_\Re\,\,|K(x-\xi,t)|\,d\xi\,\,\leq \, e^{\,-\,at}\, +\, \sqrt b\, \pi \,t \,  \, e^{\,-\,\omega \, t } \,
\eeq

\vs\no because $ I_0 ( |z|) < exp ( |z|).$ As consequence, for all $ t\,  \leq \, T, \,$  from (\ref{223})  it follows that  

\vs\beq               \label{224}
\int _0^t \, d\tau \, \int_\Re\,\,|K(x-\xi,t-\tau)|\,d\xi\,\,\leq \, ( 1  \, + \sqrt b \, \pi /\omega ) \, T 
\eeq

\vs\no while, for $t \rightarrow \infty , \, $   the formula (\ref{221}) implies

\beq               \label{225}
\int _0^t \, d\tau \, \int_\Re\,\,|K(x-\xi,t)|\,d\xi\,\,\leq \, \frac{1}{a}\, +\, \pi \sqrt b \, \int_0^\infty \, \tau \,\,e^{\,-\,\frac{\beta +a }{2}\,\tau} \,\, I_0 ( \frac{a-\beta  }{2}\,\tau) \,\,d\tau\,= 
\eeq

\[ =\, \frac{1}{a}\, +\, \pi \sqrt b \, \, \displaystyle {\frac{a+\beta}{2(a\beta)^{3/2}}}\,\,= \beta_0. \]

\vs The integrals related to the kernel $\, K_1 \, = e^{\,-\, \beta\,t }* 
\,K \, $ defined by (\ref{215}) can be estimated in the same way. As  $ | J_0 | \, \leq 1 \, $ and 

\vs\beq             \label{226}
\int_0^\infty \, E(\tau )\, d\tau \, =\, \frac{1}{a\beta}\,=\beta_1 \,,
\eeq

\vs\no the following results can be stated:

\vspace{5mm} {\bf Theorem 2.4} - {\em For all } $\, t \, \in \, (0,\infty),\,${\em \,\,the kernels }  $\,  K ,\,K_1\, $  {\em defined by } (\ref{24}), (\ref{215})  {\em  are such that }

\vs
\beq               \label{227}
\int_\Re |K| \, d\xi \leq \,  ( 1\, +\, \sqrt b \,\pi \,t \, ) \,\,e^{- \omega \, t\,}, \qquad \int_\Re |K_1| \, \ d\xi \leq \, \,E(t)\,  
\eeq
\\
\beq               \label{228}
\int_0^t\,d \tau\, \int_\Re |K| \, d\xi \leq \,  \beta_0\,,\,\qquad   \int_0^t d\tau\,  \int_\Re |K_1| \, \ d\xi \leq \, \,\beta_1\,  
\eeq

\vs \no {\em where}  $\,\omega \,= min \,( a, \beta ), \,E(t) \,$ {\em  is given by }(\ref{219}) {\em and the constants } $\, \beta _0, \, \beta_1\,${\em  are defined in } (\ref{225}), (\ref{226}).
\hbox{}\hfill\rule{1.85mm}{2.82mm}

\newpage 
\vs At last we observe that 

\beq               \label{229}
\int_\Re\,\,K_1(x-\xi,t)\,d\xi\,= \chi (t) , \qquad \int_\Re\,\,K(x-\xi,t)\,d\xi\,=  ( \partial _t + \beta ) \,\chi (t)
\eeq

\vs\no with the function $ \,\chi(t) \, $ defined by

\vs\beq               \label{230}
\chi (t)\, = \,e^{\,-\,\frac{\beta +a }{2}\,t} \,\, sen ( \varrho \,t ) / \varrho, \qquad  \varrho = \, \frac{1}{2} \, \sqrt{4b\, - ( a- \beta )^2}.
\eeq

\vs\no The formulae (\ref{229}), (\ref{230}) can be verified by means of their $\cal{L}$ transforms because the related Laplace integrals are endowed with absolute convergence.

\section{ Existence and uniqueness results  }
\label{sec:3}
  As for the data   $\,F\,  $ and $\,g\,$ of the problem $\,{\cal P}_0\,$, we shall admit:

\vs {\bf Assumption 3.1\,} {\em The function }$\, g(x) \, $ {\em is continuously differentiable and bounded together with} $\, g'(x).\,$ { \em The function }$\,F(\,x,t,u\,)\, $ {\em is defined and continuous on the set}

\beq  \label{31}
  \,\,\, D \ \equiv  \{ \, (x,t,u)\,\,: \,(x,t)  \in   \Omega_T \,, \,\,-\infty \,<\,u\,<\infty \,  \}
 \eeq

\vs\no {\em and more is  uniformly Lipschitz  continuous in }$ \,( \,x,\,t,\, u\, )\,$ {\em  for  each compact subset of }$\, \Omega_T\,. $ {\em Moreover }$\, F\, $ {\em is bounded for bounded} $\, u\,$ {\em and there exists a constant }$\,C _F\,$ {\em such that the estimate} 

\vs  
\beq \label{32}
\, |F (x,t,u_1)\,-\,F (x,t,u_2)|\, \leq \,\,C _F \,\, \ \, | u_1-u_2\,| \, 
\eeq

 \vs\no {\em holds for all }$\, (\,u_1,\,u_2\,)$. \hbox{}\hfill\rule{1.85mm}{2.82mm}

\vspace{5mm} When the problem $\,{\cal P}_0\,$ admits a solution $\, u,\,$  then $\, u\, $ must satisfy the integral equation 

 \beq                        \label{33}
   u =\int_\Re   K ( x-\xi, t) g (\xi)d\xi   + \int ^t_0     d\tau \int_\Re   K ( x-\xi, t-\tau) F[\xi,\tau, u(\xi,\tau\,)] d\xi
\eeq

\vs\no because of the properties of $ \,K\,$stated by theorem 2.2 and the assumption 3.1 on the data . On the other hand, let $\, u(x,t)\,$ be a solution of (\ref{33}) which is continuous and  bounded   so that also  $\, F(x,t,u(x,t))\,$ is  continuous and bounded  by assumption 3.1 . Then one can verify that $\,u\,$  satisfies (\ref{11}) owing to the properties of $ \, K\, , \,F \, $ and $ \,g. \,$ So, it' s possible to conclude that

\vspace{5mm}
 {\bf Theorem  3.1 - }{\em The differential  problem  }(\ref{11}) {\em  admits a unique solution only if the integral equation }(\ref{33}) {\em has a  unique  solution which is  continuous and bounded in  }$ \,\Omega_T\,$. \hbox{}\hfill\rule{1.85mm}{2.82mm}

\vspace{5mm}
Consider now the integral equation (\ref{33}) and,  for small time $\,0\,\leq t\,\leq \theta\, $ (\,with $\, \theta <\, T\,),$ let

\vspace{1mm}\beq     \label{34}
|| \, v \, ||_\theta \,  =  {\tiny {\displaystyle{\sup_ {\begin{array}{ll} 
-\infty < x <\infty  \\  \hspace{3.5mm} 0\leq  t \leq \theta  \end{array}}}}}\,| v (x,t)\,| \,\,
\eeq

\vspace{-2.5mm}
\no and

\vspace{-2mm}
\beq   \label{35}
  \,{\cal B}_ \theta \, \equiv \, \{\, v\,(\,x,t\,) : \, v\, \in  C \,\bigl[  ( -\infty, \infty) \, \times \, [\,0,\theta \,]\,\bigr]\,\, and  \, \,\,   ||v||_\theta  \, < \infty \ \}
\eeq

\vs Further, let denote by $ \,f_1 * f_2 \,$ the convolution 

\beq\label{36} 
f_1( x,t) \,* f_2(x,t)\,\,= \,\, \int_\Re \, \, f_1(\xi,t)\, \,\,\,f_2(x-\xi,t) \,\, d \xi
\eeq 

\vspace{2mm} \no   and by  $ {\cal F}v\,$
 the mapping 

\vspace{-2mm}
\beq      \label{37}
{\cal F}v\,(x,t)\, = \, K(x,t) \, * g(x)+ \int _0^t \, K(x,t-\tau)\, * F(x,\tau,v(x,\tau))\, d \tau
\eeq

\vspace{1mm} The foregoing remarks imply that $\,B_\theta\,$  is a Banach space and (\ref{37}) maps $ {\cal B}_\theta $ in  $ {\cal B}_\theta $ and  represents a continuous map for $\,0 \,\leq t\,\leq \,\theta \,$. In fact, from  (\ref{32}) and (\ref{224}) it results

\vspace{1mm} 

\beq    \label{38}
||{\cal F} v_1\,-\,{\cal F} v_2)||_\theta  \, \leq \, C_F \, \,  \theta ( \,1\,+ \,\pi \sqrt b / \omega \,) \, \,\,  ||v_1-v_2||_\theta.
\eeq

\vspace{2mm} \noindent Hence,  when one selects $ \theta$  such that $\,
 C_F\,  \,\, \theta  \,( \,1\,+ \,\pi \sqrt b/\omega \,) \,  <\, 1 \,$ 
then ${\cal F}$ is a contraction of  $ \,{\cal B}_\theta \, $ into  $ \,{\cal B}_\theta \,$ and so   has a unique fixed point $u(x,t) \in   {\cal B}_\theta $ .  

In order to show that  this result holds also  for $\, 0 \,< \, t \, \leq \, T, \, $  it suffices to proceed by induction and so the integral equation (\ref {33}) has a unique solution  for $\, 0 \, < \, t \, \leq \, ( \, n\, +\, 1\, )\, \theta , \, $ whatever the positive integer $\, n\, $ may be.

\vs {\bf Theorem  3.2 - }{\em When the data } $\, (\,F, \,g\,) $ {\em satisfy the assumption} 3.1, {\em then the initial value problem } $\,{\cal P}_0\,${\em defined in } ({\ref{11}) {\em admits a unique regular solution }$\, u(x,t)\,$ {\em in }$\, \Omega _T\,$. 
%{\em More, for this solution the following a priori estimate holds:}
\hbox{}\hfill\rule{1.85mm}{2.82mm}

\section{ A priori  estimates and continuous dependence  }
\label{sec:4}

According to the assumption 3.1, in the class of bounded solutions, let

\vs \[
\left\| g \, \right\| \, = \sup_\Re \left|g(x) \right|, \quad \left\| u \, \right\|_T \, = \sup_{\Omega_T} \left|u(x,t) \right|, \quad \left\| F \, \right\| \, = \sup_D \left|F(x,t,u) \right|.  
\]

\no The solution u(x,t) of the problem $\,{\cal P}_0\,$ depends continuosly upon the data, owing to the following theorem.

\vs
 {\bf Theorem  4.1 - }{\em Let }$\,u_1,\,u_2\,$ {\em be solutions of the problem }$\,{\cal P}_0\,$ {\em related to the data }$\, (g_1,\,F_1)\,$ {\em and} 
$\, (g_2,\,F_2)\,$  {\em which satisfy the assumption 3.1. Then there exists a positive constant }$\,C\,$ {\em such that }

\[
\left\| u_1-u_2 \, \right\|_T \, \leq \, C\, \sup_\Re \,\left|g_1-g_2 \right|  \,+\, C\,  \sup_D \,\left|F_1(x,t,u) -F_2(x,t,u)\right|,  
\]

\no {\em where} $\,C\, $ {\em depends on} $\, C_F, \, T\, $ {\em and the parameters} $\, a,\,b,\, \beta.\,$

\vspace{4mm}{\bf Proof -} The integral equations (\ref{33}) for $\,u_1\,$ and $\,u_2\,$ are differenced and the estimates of theorem 2.4 allow to apply the Gronwall lemma. \hbox{}\hfill\rule{1.85mm}{2.82mm}

\vspace{5mm}{\bf Remark 4.1 - } When the function $\, F\, $ is a known linear source  $\, f(x,t) \, $ then (\ref{33})  gives the {\em explicit solution } of the linear problem   $\,{\cal P}_0\,$ 

\beq             \label{43}
u=\, K * g \, +\, \int_0^t \, K(x,t-\tau)\,* f(x,\tau) \, d\tau.
\eeq

By means of this formula and the basic properties of the kernel K stated in theorems 2.3 and 2.4, it's possible to estimate $\, u\, $ and its derivatives together with their asymptotic behavior.\hbox{}\hfill\rule{1.85mm}{2.82mm}

\vspace{6mm}In the non linear case, the integral equation (\ref{33}) implies a priori estimates. For example:

\vs
 {\bf Theorem  4.2 - }{\em  When the data  }$\,(g,F)\,$ {\em of the non linear problem  }$\,{\cal P}_0\,$ {\em verify the assumption 3.1, then  it results: }

\beq          \label {44}
\left\| u(x,t) \, \right\|_T \, \leq \, \beta_0 \, \left \| F \right \|  \,+\,  \left \| g \right \|  \,  [ \, e^{-\,a\,t\,}\, + \sqrt b \,\pi \,\,t \,e^{-\,\omega \,t\,}]
\eeq

\vs \no {\em where the constant } $\,\beta_0\, $ {\em is defined by } (\ref{225}){\em \,\,and depends on }$\,a,\,b,\,\beta\,.$

\vs{\bf Proof -}It suffices  to apply to (\ref{33}) the estimates (\ref{223}) and (\ref{228}).\hbox{}\hfill\rule{1.85mm}{2.82mm}

\section{ Excitable models and travelling pulses  }
\label{sec:5}

As it is well known, a biologcal system is excitable if a stimulus of sufficient size can initiate a travelling pulse which will propagate through the medium. One of the most revelant examples of systems with excitable behavior is given by neural communications by nerve cells via electrical signalling \cite{i,ks}.
%{m1,m2}

Let $\, u(x,t) \,$ a transmembrane potential and let  $\, v(x,t) \,$ a variable associated with the contributions to the membrane current 
 from sodium, potassium and other ions. Then a simple example of excitable models is the Fitzhugh Nagumo system (FHN) \cite{m1,m2}:
%m1,m2
 
\vs
\beq                                                     \label{51}
  \left \{
   \begin{array}{lll}
    \displaystyle{\frac{\partial \,u }{\partial \,t }} =\,  \varepsilon \,\frac{\partial^2 \,u }{\partial \,x^2 }
     \,-\, v\,\,  + f(u ) \,  \\
\\
\displaystyle{\frac{\partial \,v }{\partial \,t } }\, = \, b\, u\,
- \beta\, v\,
\\

   \end{array}
  \right.
 \eeq

\vspace{4mm} \noindent where $\, \varepsilon \,> 0\, $  is the diffusion coifficient related to the axial current in the axon, while $\, b\,$ and $\, \beta \, $ are positive constants that characterize the model's kinetic. Further

\beq                 \label{52} 
f(u)= u\, (\, a-u \,) \,
(\,u-1\,)
 \eeq

\vs  In absence of diffusion $\, (\varepsilon\, =\,0\, ),\,$ the system (\ref{51})- (\ref{52}) is related to a two variable phase plane system with the phase portrait that varies according to different values of the parameters $\, a,\, b,\, \beta\,. $ In fact, when $\, (1-a)^2 \, < 4b/\beta, \, $ there is only the stable steady state $\, O\, $ with $ \, u\,=\,0, \, v\,=\,0.\,$ If $\, (1-a)^2 \, > 4b/\beta, \, $ there are two other steady state $ \,A \, $ and $ \,B \, $ with

\vs \[ 
u_A \, = \frac{1}{2}\, [ \,a \,+\,1-\,\sqrt{(1-a)^2\, -\, 4b/\beta} \, ] \, <\,a, \quad  v_A \, = \frac{b}{\beta}\, u_A
\]

and

\[ 
u_B \, = \frac{1}{2}\, [ \,a \,+\,1+\,\sqrt{(1-a)^2\, -\, 4b/\beta} \, ] \, <\,1, \quad  v_B \, = \frac{b}{\beta}\, u_B.
\]

\vs The state $\, B\, $ is as locally stable as $\,O\,$
while the state $\,A\,$ between them is unstable. So, a sufficiently 
strong perturbation  to a point $\, (\,\bar u , \,0)\,$ with $\, \bar u\,> u_A\,$ may induce a large phase traiectory excursion, while a small perturbation around $\, O\, \,\,\, (\,\bar u  < \,u_A)\,$  rapidly decays towards $\, O\, \,$. All this is typical of the threshold behavior.

\vspace{1.5mm} Moreover, if $\, z\,=\,x\,-\,c\,t,$ 
%{m1,m2}
travelling wave solutions of (\ref{51}) are functions $\, u(z), v(z)\,$ such that 

\vs\beq     \label {53}
\varepsilon \, \ddot u\, +\, c \dot u\, + f(u) \, - v\, =0,\,\qquad c\, \dot v \, + b\, u\, - \beta \, v\, \,=\,0.
\eeq

\vs A solitary pulse is meant to be a solution of (\ref{53}) with appropriate boundary conditions for $ \, z \rightarrow \pm \infty. \, $  A classical meaningful example \cite{h,m1,m2}is related to the particular case of $\, b\, =\,0\,$ and

\vs \beq     \label{54}
\lim_ {z\rightarrow -\infty}\,\,u(z) \,= \,1, \qquad \qquad \lim_ {z\rightarrow \infty}\,\,u(z) \,= \, \lim_ {z  \rightarrow \infty}\,\, v(z) \,=\,0.
\eeq

\vs In this case the exact analytical solution of (\ref{53}), (\ref{54}) is

\beq        \label{55}
 u(z)\, = \frac{1}{1\,+\,e^{\gamma z}}\,\,, \qquad  v(z)\, =\,0
\eeq

\vs \no  where: $ \, \gamma = \, 1/ \sqrt{2 \varepsilon \,}, \,\,\,\, \,\ c\,= \,\sqrt{\varepsilon/2} \,\,(1-2a).\,$ Several other solutions of the equation $\,(\ref{51})_1\,$ which behave like (\ref{55}) are well known in literature  \cite{bmm,cp,dr,lg,nc,smm} and are all bounded solutions that verify the assumption 3.1. As consequence, the results stated in section 3 and 4 are meaningful for the analysis of the evolution of (FHN) model in all of the space.

%cdf nella precedente nota di bibliografia non c'entra . lo ho  messo per comodita' tipografica.
% ks e' gia' stato ripetuto altrove ma serve perche' contiene soluzioni esplicite

\vs Let 

%\vspace{2mm}
\beq      \label{56}
u(x,0)\, =\,u_0 \,, \qquad v(x,0)\, =\,v_0  \qquad \qquad ( \,x\,  \in \Re\,)
\eeq

\vspace{4mm} \no the initial conditions related to the system (\ref{51}) and let

\vs\beq      \label{57}
f(u)\, =\,-\,a\,u\, +\,\varphi(u) \quad with \quad  \varphi \,=\, u^2\, (\,a+1\,-u\,).
\eeq

\vs
\vs\no As  $(\ref{51})_2$  implies

\beq      \label{58}
v\, =\,v_0 \, e^{\,-\,\beta\,t\,} \,+\, b\, \int_0^t\, e^{\,-\,\beta\,(\,t-\tau\,)}\,u(x,\tau) \, d\tau,
\eeq

\vs\no when one puts $ \,\, F(x,t,u)\, =\,\varphi (u) \, -\, v_0(x) \, e^{\,-\,\beta\,t\,},\,\,$ then the initial value problem (\ref{51})-(\ref{56}) can be given the form (\ref{11}) to obtain $\, u\, $ in terms of the data.

\vspace{4.5mm} For this we apply the formula (\ref{33}) with $\, g\,= u_0 \,$ and put

\vs\[
K  \otimes F\,=\, \int_0^t\,d\tau\, \int_\Re \,K(x-\xi,t-\tau) \, \,F \,[\,\xi, \tau ,u(\xi,\tau)\,]\,d\xi.
\]

\vspace{4mm} \no Because of (\ref{218}) it is $\, K \, \otimes (v_0 \, e^{\,-\, \beta \,t \,}) \, = \, v_0 * K_1,\,$ and so

\vspace{4mm}\[
K\otimes F \, =\, K\otimes \varphi \,-\,\,K_1 *\, v_0 (x).
\] 

\vs\no Therefore by (\ref{33}) we get:

\vs\beq           \label{59} 
u(x,t) \,=\, u_0 * K \,-\, v_0 * K_1\, +\, \varphi \otimes K
\eeq

\vs\no and this formula, together with (\ref{58}), allows to obtain also  $\, v(x,t) \,$ in terms of the data.  If we observe that

\newpage
\vs\[
 \int_0^t\, e^{\,-\,\beta\,(\,t-\tau\,)}\,K_1(x,\tau) \, d\tau \,\,= \,K_2 (x,t) \, =\,\]
\\
\beq      \label{510}
=\, \int_0^t\,\frac{e^{- \frac{x^2}{4\varepsilon y}\,- a\,y\,-\,\beta(t-y)}}{2\,\sqrt{\,\pi\,\varepsilon \,y}} \,\, \,\ \sqrt{\frac{t-y}{b\,y}} \,\,\,\, J_1 (\,2 \,\sqrt{\,b\,y\,(t-y)\,}\,\,)\,\,dy\,\,
\eeq

\vspace{4mm}\no by means of  (\ref{58})-(\ref{510}) we get

\vs\beq       \label{511}
v(x,t) \,=\,v_0\, e^{\,-\beta\,t} \ +\, b\,[ \,u_0 * K_1\, -\, v_0*K_2\,+\,\,\varphi \otimes K_1 \,].
\eeq

\vs \no When the non linear part $\, \varphi(u)\, $ of  $\,f(u)\,$is neglegible, or can be approximated by a constant $\, \varphi_0,\, $ then  
(\ref{59})-(\ref{511})  represent the explicit solution $\, (u,v)\,$ of the linear case. Otherwise (\ref{59}) is an integral equation for the unknown $\, u\,$ that is identical to (\ref{33})    and then (\ref{511}) can be applied for $\,v.\,$ 

Because of theorems 3.2 and 4.1, the following result can be stated.

\vspace{4mm}
 {\bf Theorem  5.1 - }{\em When the data }$\, (u_0, v_0)\,${\em  satisfy the hypotheses 3.1, then the initial value problem related to (FHN) system} 
(\ref{51}) {\em has a unique regular solution in the space of bounded solution. Moreover this solution depends continuously upon the data.}\hbox{}\hfill\rule{1.85mm}{2.82mm}

\vs In order to apply theorem 4.2, an estimate like  (\ref{227}) must be set also for the kernel $\, K_2.\, $ From  (\ref{510}) it results:

\beq   \label{512}
\int_\Re \left|K_2 (x-\xi,t)\right| \, d\xi \, \leq \, \int_0^t \, \, e^{\,-\,ay\, - \beta(t-y)\,}\,(\,t-y\,) \,dy \,\leq \, t\, E(t)
\eeq

\vs\no so that, if 

\[\quad \left\| u_0 \, \right\| \, = \sup_\Re \left|u_0(x) \right|,  
 \quad \left\| v_0 \, \right\| \, = \sup_\Re \left|v_0(x) \right|, \quad 
\left\| \varphi \, \right\| \, = \sup_D \left|\varphi(u) \right|,
\]

\vs\no by means of (\ref{512}) and theorem 2.4 the formulae (\ref{59}) - (\ref{511}) imply the following conclusion.

\vspace{3mm}
 {\bf Theorem  5.2 - }{\em For initial data }$\, (u_0, v_0)\,$ {\em  compatible with assumption  3.1, the solution of the problem } 
(\ref{51}) - (\ref{56}) {\em satisfies the estimates:}

\vs 
\beq            \label{513}
\left\{ 
 \begin{array}{lll}                                                   \label{513}
 \left| u \, \right| \, \leq  \left\| u_0 \right\| \, (1+\pi \sqrt b \, t ) \, e^ {\,-\omega\,t\,}\,+\,\left\| v_0 \right\|\,E(t) \, +\, \beta_0 \,\left\| \varphi \right\|\, 
   \\
\\
\left| v \, \right| \, \leq  \left\| v_0 \right\|\, e^ {\,-\,\beta\,t\,}\,+\,b\,(\,\left\| u_0 \right\|\,+\, t\, \left\| v_0 \right\|\,) \, E(t) \, + \, b\, \beta_1\, \left\| \varphi \right\| \,
\\ 
   \end{array}
  \right.
 \eeq

\vs \no {\em with }$\,\beta_0,\, \beta_1,\, E(t)\,$  {\em defined by} (\ref{225}), (\ref{226}), (\ref{219}).  
\hbox{}\hfill\rule{1.85mm}{2.82mm}

\vs  The analysis and the stability of solutions of nonlinear binary reaction - diffusion systems of PDE's, as well as the existence of global compact attractors, have been discussed in a great number of recent and interesting papers (see e. g. \cite{chs,lw,r1,r2,sm}).  As it is  well known,  the (FHN) system admits arbitrary large  invariant rectangles  $\, \Sigma\,$  containing $\,(0,0)  \,$ so that the solution $\,(u,v),\, $ for all times $\, t\,>\,0,\,$ lies in the interior of $\, \Sigma\,$ when the initial data $\,(u_o,v_o)\,$  belong to $\, \Sigma\,.$  Besides, when both $\,\beta\,$ and $\,a\,$ are strictly positive, then the estimates (\ref{513}) confirm that the large time behaviour of the solution is determined only by the reaction mechanism because the explicit terms depending on the initial data are exponentially vanishing.

Finally we observe that  when  $\, b\,= \,0,\,\beta\, =\, 0\, $ there is no bounded set that attracts all solutions  and the asymptotic analysis needs the use of characteristic parameters such as fast and slow times, together with  rigorous estimates uniformly valid for all $\, t.\,$ To this extent, the basic properties of the kernel $\, K  $ might be useful for a rigorous singular perturbation analysis \cite{h,k,m2}. 

\begin{acknowledgements}
This work has been performed under the auspices of the G.N.F.M. of   I.N.D.A.M. and M.I.U.R. (P.R.I.N. 2007) " Waves and stability in continuous media ".We thank professor Rionero for his suggestions.
\end {acknowledgements}

\end{document}